\documentclass[usenatbib,useAMS]{mn2e}
\usepackage{graphics}
\usepackage{psfrag}
\usepackage{xspace}
\usepackage{amstext}
\usepackage{amsmath}
\usepackage{amssymb}
\usepackage{natbib}
\usepackage{aas_macros}
\usepackage{ulem}
\usepackage{graphicx}
\usepackage{longtable}
\usepackage{rotating}
\usepackage{float}
\newcommand{\FigDir}[1]{./#1}

\newcommand{\yrs}{{\rm years}\xspace}

\newcommand{\mnths}{{\rm months}\xspace}
\newcommand{\days}{{\rm days}\xspace}

\newcommand{\en}{OGLE-BLG182.1.162852\xspace}

\newlength{\voff}
\newcommand{\ogleD}{OGLE-LMC-ECL-11893\xspace}
\newcommand{\ogleG}{OGLE-LMC-ECL-17782\xspace}
\newcommand{\eeCeph}{EE~Cephei\xspace}
\newcommand{\epsAur}{\ensuremath{\varepsilon} Aurigae\xspace}
\newcommand{\kh}{KH 15D\xspace}
\newcommand{\Jfourteen}{J1407\xspace}

\begin{document}

\title[\en: An Eclipsing Binary with a Circumstellar Disk]{\en: An Eclipsing Binary with a Circumstellar Disk}

\author[Rattenbury, N.~J.]
{N.~J.~Rattenbury$^{1}$\thanks{e-mail: n.rattenbury@auckland.ac.nz},
{\L}. Wyrzykowski$^{2,3}$,
Z. Kostrzewa-Rutkowska$^{2}$,\newauthor
A. Udalski$^{2}$, 
S. Koz{\l}owski$^{2}$,
M. K. Szyma{\'n}ski$^{2}$,
G. Pietrzy{\'n}ski$^{2,4}$,
I. Soszy{\'n}ski$^{2}$,\newauthor
R. Poleski$^{2,5}$,
K. Ulaczyk$^{2}$,
J. Skowron$^{2}$,
P. Pietrukowicz$^{2}$,
P. Mr{\'o}z$^{2}$,\newauthor
D. Skowron$^{2}$\\
$^1$Department of Physics, University of Auckland, Private Bag 92019, Auckland, New Zealand\\
$^2$Warsaw University Observatory, Al.~Ujazdowskie 4, 00-478 Warszawa, Poland\\
$^3$Institute of Astronomy, University of Cambridge, Madingley Road, CB3~0HA Cambridge, UK\\
$^4$Universidad de Concepci{\'o}n, Departamento de Astronomia, Casilla 160--C, Concepci{\'o}n, Chile\\
$^5$Department of Astronomy, Ohio State University, 140 W. 18th Ave., Columbus, OH 43210, USA
}
\date{Accepted ........
      Received .......;
      in original form ......}

\pubyear{2014}

\maketitle
\begin{abstract}
We present the discovery of a plausible  disk-eclipse system \en. The OGLE light curve for \en shows three episodes of dimming by I $\simeq 2 - 3$ magnitudes, separated by 1277 days. The shape of the light curve during dimming events is very similar to that of known disk eclipse system \ogleD \citep{2014ApJ...788...41D}. The event is presently undergoing a dimming event, predicted to end on December 30th, 2014. We encourage spectroscopic and multi-band photometric observations now. The next dimming episode for \en is expected to occur in March 2018. 
\end{abstract}


\section{Introduction}
\label{sec:intro}
There has been a recent increased interest in discovering astronomical events which show transient intervals of dimming (see e.g. \citealt{2014MNRAS.441.2691Q, 2014MNRAS.441.3733M}). Analogous to events which show a transient brightening, such as novae or microlensing, dimming events also offer the opportunity to investigate interesting astrophysical phenomena. 

The large-scale long-term sky variability survey OGLE \citep{2003AcA....53..291U}  routinely observes millions of stars towards the Galactic bulge and the Magellanic Clouds. The OGLE database is a rich resource for the discovery of a wide range of transient phenomena. In an initial analysis of data from the last three phases of the OGLE project (OGLE-II, OGLE-III and OGLE-IV), we have discovered a number of events which show instances of transient dimming. The full results of this analysis will be presented in a later paper. Briefly; in our analysis we computed the von Neumann statistic \citep{vonNeumann, 2014ApJ...781...35P} and skew for each OGLE light curve and investigated those light curves having extreme values for each of these two statistics. One such light curve, for star \en (17\wh59\wm3\fs54, -30\wdg49\wm6\fs0), shows three deep, asymmetric dimming events.

\section{Light curve properties}
\begin{figure}
\includegraphics[width=84mm]{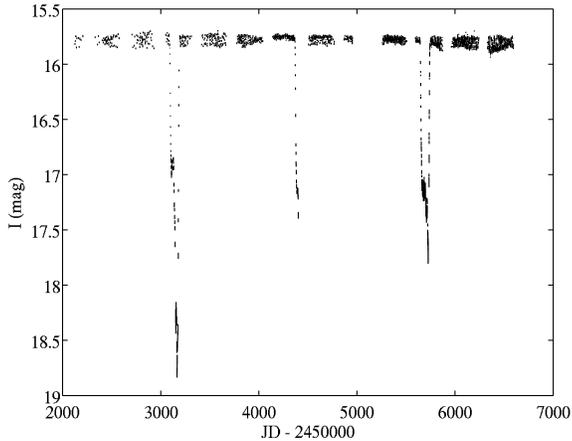}
 \caption{\label{fig:fullLightcurve}OGLE-III and OGLE-IV (after epoch 2455000) light curve data for event \en.}
  \end{figure}

\begin{figure}
\includegraphics[width=84mm]{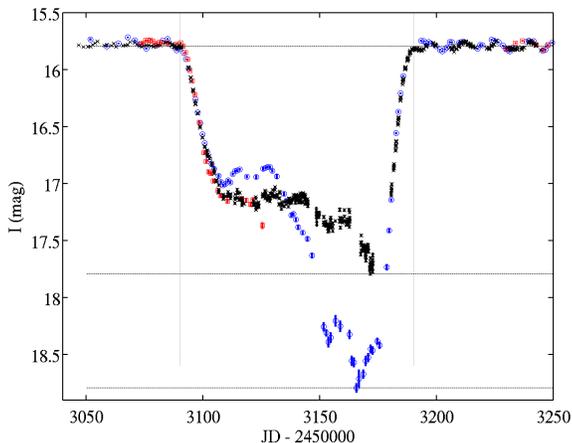}
 \caption{\label{fig:PhasedEclipses}The three transient dimming events for \en. The data corresponding to the first, second and third dimming events are shown with blue circles, red squares and black crosses respectively. A colour version of this plot is available in the online publication of this paper.}
  \end{figure}

The OGLE light curve for \en is shown in Figure~\ref{fig:fullLightcurve}. The three deep minima are spaced by $\sim 1277$ days (3.5 years).
Figure~\ref{fig:PhasedEclipses} shows the data for the second and third dimming events overlaid on the first dimming event. The light curve is well sampled during ingress for all three dimming instances while egress was only covered for the first and third dimming events. While we do not have data which extends completely through the second dimming event, the shape of the light curve during the second dimming event is largely consistent with that of the first and third dimming events. The interval between dimming events is constant, at 1277 days. We speculate that whatever the physical cause of these transient dimming events for \en, they occur with a period of 1277 days. 
The light curve profile for \en shows a striking similarity to that for \ogleD, see Figure~1 of \cite{2014ApJ...788...41D}. \cite{2014ApJ...788...41D} find a close similarity between the light curve of \ogleD and \eeCeph, an eclipsing binary with a circumstellar disk \citep{2012AA...544A..53G} and suggest that \ogleD is also a disk-eclipse system. We suggest, in turn, that star \en similarly is an eclipsing binary star with a circumstellar disk around one of the binary star components, the disk periodically obscuring the light from the other star in the binary system.

The depths of the first and third eclipses are $\Delta I \simeq 3$ mag and $\Delta I \simeq 2$ mag respectively. The second eclipse event was not fully covered by OGLE. 
The eclipse
duration is $~\simeq 100$ days, with the ingress/egress lasting $~\simeq 10$ days, and
the declining trough lasting $~\simeq 80$ days.

\begin{figure}
\includegraphics[width=90mm]{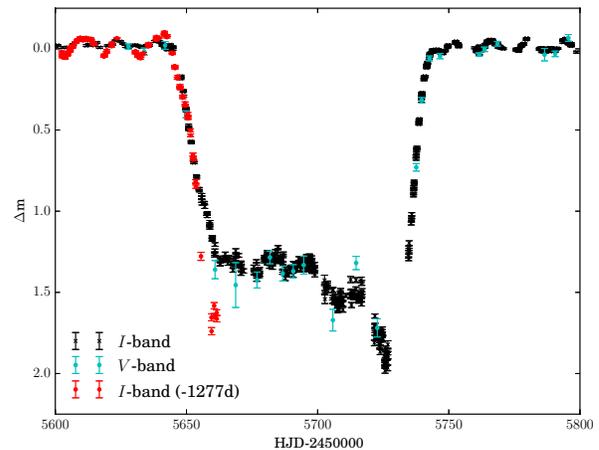}
 \caption{\label{fig:OGLEVandI}Magnitude change with respect to baseline for star \en in both $V$- (cyan points) and $I$-bands (black points), for the third dimming episode shown in Figure~\ref{fig:fullLightcurve}. Red points indicate the $I$-band data for a dimming event in progress at the time of writing (OGLE-IV). A colour version of this plot is available in the online publication of this paper.}
  \end{figure}

Some $V$-band data were collected for \en during the third dimming event. The difference between baseline magnitude and the magnitude during the dimming event, $\Delta m$, is shown in Figure~\ref{fig:OGLEVandI}, for both $V$- and $I$-bands. Since $\Delta m_I \simeq \Delta m_V$, this particular dimming event was colourless. Also shown in Figure~\ref{fig:OGLEVandI}, are the OGLE $I$-band data available at the time of writing. Star \en is presently undergoing another dimming event. Data for the present dimming event initially followed the same ingress profile as for previous dimming episodes, but now appear to be tracing out a deeper minimum than that seen previously.

The baseline magnitude for \en is clearly variable. A spectral analysis of the out-of-eclipse data gives a dominant period of 14.37 days. The average amplitude of the 14\fd37 period signal in the baseline data is $\sim 0.06$ mag, however, there is clearly variation in the amplitude, and there may be other frequency components present. More detailed analysis of the short period variability of star \en will be presented in a later paper. 

\section*{Discussion}
Only a handful of disk eclipse systems are known. The basic observed parameters of known disk eclipse systems are given in Table~\ref{tab:diskEclipseSystems}, along with those for star \en. 

\begin{table*}
\begin{minipage}{150mm}
  \caption{\label{tab:diskEclipseSystems}Basic observable parameters of known disk-eclipse binary stars, and for star \en.}
  \begin{tabular}{@{}lllll@{}}
  \hline
  System & Period & Eclipse Duration & Eclipse Depth & Reference(s) \\
  \hline
  \epsAur & 27.1 \yrs & $\sim22$ \mnths & 0.9 (V) & \cite{2010Natur.464..870K, 2010AJ....139.1254S}\\
  \eeCeph & 5.6 \yrs & $\sim 30 - 90$ \days & 0.5 -- 2.0 (B,V) &  \cite{2012AA...544A..53G,2003AA...403.1089G}\\
  \kh & 48.4 \days & variable\footnote{presently $\sim 24$ \days} & $\sim 4$ (I) & \cite{2005AJ....130.1896H,2006ApJ...644..510W} \\
  \Jfourteen & $> 850$ \days & $\sim 54$\footnote{Duration of the single eclipse event observed to date, during which multiple dimming events occurred.} \days & $> 3.3$ (V) & \cite{2012AJ....143...72M} \\
  \ogleG & 13.35 \days & 2.67 \days & $\sim 0.4$ (I) & \cite{2011AcA....61..103G}\\
  \ogleD & 468.124 \days & $\sim 15$ \days & 1.5 (I) & \cite{2014ApJ...788...41D}\\
 \textbf{\en}& \textbf{1277 days} & \textbf{100 days }& \textbf{2 -- 3 (I)} & \textbf{This work}\\
   \hline
\end{tabular}
\end{minipage}
\end{table*}

The values for the period, duration and depth of the dimming events in star \en are broadly consistent with those attributed to transits of the disk in the known disk-eclispe systems listed in Table~\ref{tab:diskEclipseSystems}.

The eclipse depth for other known disk-eclipse events \epsAur, \ogleG and \kh vary from one eclipse to another and this is considered to be owing to the precession of the disks in these systems. Figure~\ref{fig:PhasedEclipses} shows that the depth of the dimming events for \en also changes from event to event, suggesting that, given a circumstellar disk is the cause of the dimming events, the disk is precessing. 

A counterpart to star \en was found in the GLIMPSE database \citep{2003PASP..115..953B} of {\it Spitzer} photometry \citep{2004ApJS..154....1W}, and also in the 2MASS database \citep{2006AJ....131.1163S}. Star G359.8611-03.4652 (2MASS 17590354-3049059) lies 0.124 arcsec from the OGLE co-ordinates for star \en and has infrared magnitudes as given in Table~\ref{tab:GlimpseTwoMASS}. Assuming a constant period for the dimming episodes, the Spitzer and 2MASS observations were taken at out-of-eclipse times for \en.  The {\it Spitzer} colours for \en are consistent with a bare stellar atmosphere \citep{2004ApJS..154..367M}. There is no infra-red excess which could be associated with a circumstellar disk or envelope of gas or dust. It is worth noting that \cite{2012AJ....143...72M} similarly found no infra-red excess in the 2MASS photometry associated for disk-eclipse event \Jfourteen.

\begin{table}
\caption{\label{tab:GlimpseTwoMASS}Infra-red photometry for the counterpart to \en in the {\it Spitzer}/GLIMPSE and 2MASS databases.}
\begin{tabular}{|ccc|cccc|}
\hline
\multicolumn{7}{|c|}{\en}\\
\hline
\multicolumn{3}{|c|}{2MASS} & \multicolumn{4}{c|}{GLIMPSE}\\
J & H & K & 3.6 & 4.5 & 5.8 & 8.0\\
\hline
13.778 & 12.834 & 12.545 & 12.317 & 12.346 & 12.426 & 12.154 \\
\hline
\end{tabular}
\end{table}

\begin{figure}
\includegraphics[width=84mm]{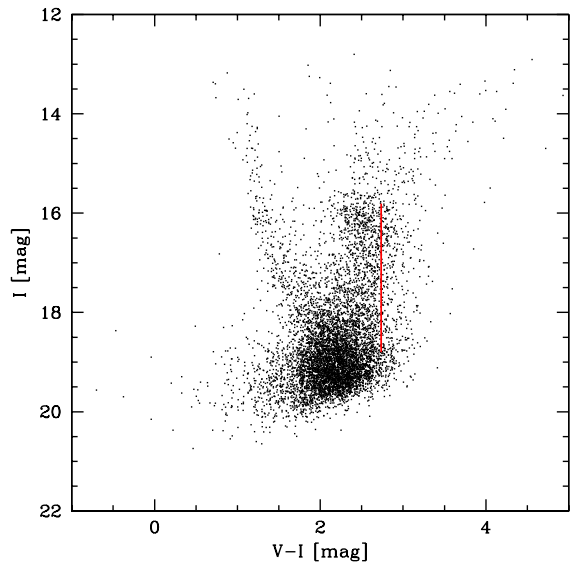}
 \caption{\label{fig:cmd}Colour-magnitude diagram of OGLE field stars for star \en. The position on the CMD for star \en is represented by the straight red line, for times both in, and out of, dimming episodes.}
  \end{figure}

The position of star \en on a colour-magnitude diagram of OGLE field stars is shown in Figure~\ref{fig:cmd}, for both in- and out-of-eclipse times. Out-of-eclipse, star \en has a magnitude and colour consistent with a K giant star. 

The small amplitude variations in the out-of-eclipse data for \en are consistent with the presence of spots on the star \citep{2013AcA....63...21S}. The variations in the baseline have a similar amplitude and period to the 9.6 day, 0.1 mag quasi-periodic variations for \kh \citep{2005AJ....130.1896H}. The visible component of the \kh system is a weak-lined T-Tauri star and are attributed to star spots on the surface of this star. \Jfourteen shows 0.1 mag out-of-eclipse variations as well, with a somewhat shorter period of 3.2 days. These variations, too, are consistent with the presence of star spots \citep{2012AJ....143...72M}. 

Without spectroscopic measurements of the visible component of \en we can only speculate on what type of star it is. Similarly, spectroscopic observations during a dimming episode would allow firmer conclusions on what is causing the periodic dimming events. 

\section*{Conclusion}
The OGLE light curve for \en is strikingly similar to light curves for binary star systems comprising an eclipsing disk around one of the binary star members. Baseline variability is also consistent with that seen in the light curves of the visible components of other binary disk eclipsing events.  The event is presently undergoing a dimming event, which will end around December 30th, 2014. The next dimming episode will be in March 2018. Spectroscopic observations of \en are strongly encouraged now.

\section*{Acknowledgements}
The OGLE project has received funding from the European Research Council under the European Community's Seventh Framework Programme (FP7/2007-2013)/ERC grant agreement no. 246678 to AU.
This work has been supported by the Polish National Science Centre grant No. DEC-2011/03/B/ST9/02573 to IS.

This publication makes use of data products from the Two Micron All Sky Survey, which is a joint project of the University of Massachusetts and the Infrared Processing and Analysis Center/California Institute of Technology, funded by the National Aeronautics and Space Administration and the National Science Foundation. This work is based in part on observations made with the {\it Spitzer Space Telescope}, which is operated by the Jet Propulsion Laboratory, California Institute of Technology under a contract with NASA. This research has made use of the SIMBAD database,
operated at CDS, Strasbourg, France. NJR is a Royal Society of New Zealand Rutherford Discovery Fellow.

\bibliographystyle{latest}

\bibliography{microlensing,eclipsingBinaries,surveys}

\begin{thebibliography}{19}
\providecommand{\natexlab}[1]{#1}
\providecommand{\url}[1]{\texttt{#1}}
\providecommand{\urlprefix}{URL }

\bibitem[{{Benjamin} et~al.(2003){Benjamin}, {Churchwell}, {Babler}, {Bania},
  {Clemens}, {Cohen}, {Dickey}, {Indebetouw} et~al.}]{2003PASP..115..953B}
{Benjamin}, R.A., {Churchwell}, E., {Babler}, B.L., {Bania}, T.M., {Clemens},
  D.P., {Cohen}, M., {Dickey}, J.M., {Indebetouw}, R., et~al., 2003, \pasp,
  115, 953

\bibitem[{{Dong} et~al.(2014){Dong}, {Katz}, {Prieto}, {Udalski}, {Kozlowski},
  {Street}, {Bramich}, {Tsapras} et~al.}]{2014ApJ...788...41D}
{Dong}, S., {Katz}, B., {Prieto}, J.L., {Udalski}, A., {Kozlowski}, S.,
  {Street}, R.A., {Bramich}, D.M., {Tsapras}, Y., et~al., 2014, \apj, 788, 41

\bibitem[{{Ga{\l}an} et~al.(2012){Ga{\l}an}, {Miko{\l}ajewski}, {Tomov},
  {Graczyk}, {Apostolovska}, {Barzova}, {Bellas-Velidis}, {Bilkina}
  et~al.}]{2012AA...544A..53G}
{Ga{\l}an}, C., {Miko{\l}ajewski}, M., {Tomov}, T., {Graczyk}, D.,
  {Apostolovska}, G., {Barzova}, I., {Bellas-Velidis}, I., {Bilkina}, B.,
  et~al., 2012, \aap, 544, A53

\bibitem[{{Graczyk} et~al.(2003){Graczyk}, {Miko{\l}ajewski}, {Tomov}, {Kolev},
  and {Iliev}}]{2003AA...403.1089G}
{Graczyk}, D., {Miko{\l}ajewski}, M., {Tomov}, T., {Kolev}, D., {Iliev}, I.,
  2003, \aap, 403, 1089

\bibitem[{{Graczyk} et~al.(2011){Graczyk}, {Soszy{\'n}ski}, {Poleski},
  {Pietrzy{\'n}ski}, {Udalski}, {Szyma{\'n}ski}, {Kubiak}, {Wyrzykowski}
  et~al.}]{2011AcA....61..103G}
{Graczyk}, D., {Soszy{\'n}ski}, I., {Poleski}, R., {Pietrzy{\'n}ski}, G.,
  {Udalski}, A., {Szyma{\'n}ski}, M.K., {Kubiak}, M., {Wyrzykowski}, {\L}.,
  et~al., 2011, \actaa, 61, 103

\bibitem[{{Hamilton} et~al.(2005){Hamilton}, {Herbst}, {Vrba}, {Ibrahimov},
  {Mundt}, {Bailer-Jones}, {Filippenko}, {Li} et~al.}]{2005AJ....130.1896H}
{Hamilton}, C.M., {Herbst}, W., {Vrba}, F.J., {Ibrahimov}, M.A., {Mundt}, R.,
  {Bailer-Jones}, C.A.L., {Filippenko}, A.V., {Li}, W., et~al., 2005, \aj, 130,
  1896

\bibitem[{{Kloppenborg} et~al.(2010){Kloppenborg}, {Stencel}, {Monnier},
  {Schaefer}, {Zhao}, {Baron}, {McAlister}, {ten Brummelaar}
  et~al.}]{2010Natur.464..870K}
{Kloppenborg}, B., {Stencel}, R., {Monnier}, J.D., {Schaefer}, G., {Zhao}, M.,
  {Baron}, F., {McAlister}, H., {ten Brummelaar}, T., et~al., 2010, \nat, 464,
  870

\bibitem[{{Mamajek} et~al.(2012){Mamajek}, {Quillen}, {Pecaut}, {Moolekamp},
  {Scott}, {Kenworthy}, {Collier Cameron}, and {Parley}}]{2012AJ....143...72M}
{Mamajek}, E.E., {Quillen}, A.C., {Pecaut}, M.J., {Moolekamp}, F., {Scott},
  E.L., {Kenworthy}, M.A., {Collier Cameron}, A., {Parley}, N.R., 2012, \aj,
  143, 72

\bibitem[{{Megeath} et~al.(2004){Megeath}, {Allen}, {Gutermuth}, {Pipher},
  {Myers}, {Calvet}, {Hartmann}, {Muzerolle} et~al.}]{2004ApJS..154..367M}
{Megeath}, S.T., {Allen}, L.E., {Gutermuth}, R.A., {Pipher}, J.L., {Myers},
  P.C., {Calvet}, N., {Hartmann}, L., {Muzerolle}, J., et~al., 2004, \apjs,
  154, 367

\bibitem[{{Meng} et~al.(2014){Meng}, {Quillen}, {Bell}, {Mamajek}, {Scott}, and
  {Zhou}}]{2014MNRAS.441.3733M}
{Meng}, Z., {Quillen}, A.C., {Bell}, C.P.M., {Mamajek}, E.E., {Scott}, E.L.,
  {Zhou}, J.L., 2014, \mnras, 441, 3733

\bibitem[{{Price-Whelan} et~al.(2014){Price-Whelan}, {Ag{\"u}eros}, {Fournier},
  {Street}, {Ofek}, {Covey}, {Levitan}, {Laher} et~al.}]{2014ApJ...781...35P}
{Price-Whelan}, A.M., {Ag{\"u}eros}, M.A., {Fournier}, A.P., {Street}, R.,
  {Ofek}, E.O., {Covey}, K.R., {Levitan}, D., {Laher}, R.R., et~al., 2014,
  \apj, 781, 35

\bibitem[{{Quillen} et~al.(2014){Quillen}, {Ciocca}, {Carlin}, {Bell}, and
  {Meng}}]{2014MNRAS.441.2691Q}
{Quillen}, A.C., {Ciocca}, M., {Carlin}, J.L., {Bell}, C.P.M., {Meng}, Z.,
  2014, \mnras, 441, 2691

\bibitem[{{Skrutskie} et~al.(2006){Skrutskie}, {Cutri}, {Stiening}, {Weinberg},
  {Schneider}, {Carpenter}, {Beichman}, {Capps} et~al.}]{2006AJ....131.1163S}
{Skrutskie}, M.F., {Cutri}, R.M., {Stiening}, R., {Weinberg}, M.D.,
  {Schneider}, S., {Carpenter}, J.M., {Beichman}, C., {Capps}, R., et~al.,
  2006, \aj, 131, 1163

\bibitem[{{Soszy{\'n}ski} et~al.(2013){Soszy{\'n}ski}, {Udalski},
  {Szyma{\'n}ski}, {Kubiak}, {Pietrzy{\'n}ski}, {Wyrzykowski}, {Ulaczyk},
  {Poleski} et~al.}]{2013AcA....63...21S}
{Soszy{\'n}ski}, I., {Udalski}, A., {Szyma{\'n}ski}, M.K., {Kubiak}, M.,
  {Pietrzy{\'n}ski}, G., {Wyrzykowski}, {\L}., {Ulaczyk}, K., {Poleski}, R.,
  et~al., 2013, \actaa, 63, 21

\bibitem[{{Stefanik} et~al.(2010){Stefanik}, {Torres}, {Lovegrove}, {Pera},
  {Latham}, {Zajac}, and {Mazeh}}]{2010AJ....139.1254S}
{Stefanik}, R.P., {Torres}, G., {Lovegrove}, J., {Pera}, V.E., {Latham}, D.W.,
  {Zajac}, J., {Mazeh}, T., 2010, \aj, 139, 1254

\bibitem[{{Udalski}(2003)}]{2003AcA....53..291U}
{Udalski}, A., 2003, Acta Astronomica, 53, 291

\bibitem[{{von Neumann} et~al.(1941){von Neumann}, {Kent}, {Bellinson}, and
  {Hart}}]{vonNeumann}
{von Neumann}, J., {Kent}, R.H., {Bellinson}, H.R., {Hart}, B.I., 1941, The
  Annals of Mathematical Statistics, 12, 153

\bibitem[{{Werner} et~al.(2004){Werner}, {Roellig}, {Low}, {Rieke}, {Rieke},
  {Hoffmann}, {Young}, {Houck} et~al.}]{2004ApJS..154....1W}
{Werner}, M.W., {Roellig}, T.L., {Low}, F.J., {Rieke}, G.H., {Rieke}, M.,
  {Hoffmann}, W.F., {Young}, E., {Houck}, J.R., et~al., 2004, \apjs, 154, 1

\bibitem[{{Winn} et~al.(2006){Winn}, {Hamilton}, {Herbst}, {Hoffman}, {Holman},
  {Johnson}, and {Kuchner}}]{2006ApJ...644..510W}
{Winn}, J.N., {Hamilton}, C.M., {Herbst}, W.J., {Hoffman}, J.L., {Holman},
  M.J., {Johnson}, J.A., {Kuchner}, M.J., 2006, \apj, 644, 510

\end{thebibliography}

\end{document}